# PERFORMANCE AND COMPLEXITY ANALYSIS OF A REDUCED ITERATIONS LLL ALGORITHM

Nizar OUNI[1] and Ridha BOUALLEGUE[2]

[1]National Engineering School of Tunis, SUP'COM, InnovCom laboratory, Tunisia
[2]SUP'COM, InnovCom laboratory, Tunisia

## ABSTRACT

*Multiple-input multiple-output (MIMO) systems are playing an increasing and interesting role in the recent wireless communication. The complexity and the performance of the systems are driving the different studies and researches. Lattices Reduction techniques bring more resources to investigate the complexity and performances of such systems.*

*In this paper, we look to modify a fixed complexity verity of the LLL algorithm to reduce the computation operations by reducing the number of iterations without important performance degradation. Our proposal shows that we can achieve a good performance results while avoiding extra iteration that doesn't bring much performance.*

## 1. INTRODUCTION

MIMO communication systems are introduced to combat fading and provide high data rate. The MIMO system consists of transmitting multiple independent data symbols via multiple antennas. For the reception, a MIMO decoder needs to be used to detect, estimate, and reconstruct the received symbols. Multiple detection schemes can be used, such as the zero-forcing (ZF) or the minimum mean square error (MMSE) criterion. Also, the maximum likelihood decoder (ML) is considered as the optimal solution for the MIMO detection in term of Bit Error Rate (BER). But, unfortunately the ML algorithm seems to be complex for hardware implementations. Therefore, linear MIMO detection techniques like ZF and MMSE are better in term of complexity, but suffer from BER performance degradation.

The lattice-reduction (LR) preprocessing technique has been proposed to be used with linear detection in order to transform the system model into an equivalent system with better channel matrix's effect and so to reduce the complexity of the system. It was shown in previous studies that LR techniques improve the BER performances significantly.

The populated LR algorithm is called Lenstra-Lenstra-Lovàsz (LLL) algorithm is the most used one. It was called according to the name of the inventors [1]. But, the LLL algorithm brings many challenges due to higher processing complexity and the undeterministic execution time [2].

LLL algorithm has a major limit which is the varying complexity that could be large and limits the decoding speed of the communication system. But, it is always presenting the best performance in term BER. The complex Lenstra-Lenstra-Lovàsz algorithm (CLLL) [3]is applying the basis reduction for complex field, while the LLL is targetinga real valued matrix. The different studies and simulation results show that CLLL requires less processing operations [4]. Effective LLL algorithm (ELLL) [5], come with a new idea that consists to change the Lovàsz reduction condition in order to relax the related equations. Also, the FcLLL prposed by Wen [2] reduces the number of iterations for the algorithm to fix iteration number instead of infinite iterations. This technique improves the complexity but remains worse than LLL in term of BER performance.

In this paper we, will focus on the FcLLL algorithm using ZF decoding technique and we propose some modifications to the original FcLLL algorithm to keep a reduced number of loops and targeting a good BER.

## 2. SYSTEM MODEL DESCRIPTION

During this paper we will consider that $(.)^H$ and $(.)^T$ denote respectively the hermission transpose and the transpose of a matrix.

We consider the spatial multiplexing MIMO system with $N_t$ transmit and $N_r$ receive antennas with a Rayleigh channel non variant in the time.

$$x = H.s + n \tag{1}$$

Where $s = [s_1, s_2, \ldots, s_{N_t}]^H$; $(s_i \in s)$ is the information vector with $S$ being a constellation set of square QAM with $E[ss^H] = \sigma_s^2.I_{N_t}$ and the real and imaginary parts are $\{-\sqrt{M_S} + 1, \ldots, -1, 1, \ldots, \sqrt{M_S} - 1\}$ with $M_S$ being the constellation size, . We will suppose that the average transmit power of each antenna is normalized to one, so $E[ss^H] = I_{N_t}$. With $I_m$ is the m × m identity matrix.

$H$is an $N_r \times N_t$; $(N_r \geq N_t)$complex channel matrix, $x = [x_1, x_2, \ldots, x]^T$is the received signal vector, and $n = [n_1, n_2, \ldots, n_{N_r}]^T$is the complex additive white Gaussian noise (AWGN) vector with zero mean and covariance $\sigma_n^2.I_{N_r}$.

On the receiver side, $x = [x_1, x_2, \ldots, x_{N_r}]^T$ are the symbols at receiver's respective antennas which will be used to estimate transmitted e the symbols [4]. The receiver will analyze all received information to compute the transmitted data. So, a detection, computation, equalization and estimation of the received data will happen.

At receiver side, the linear zero forcing (ZF) detector compute the inverse of the channel matrix to estimate the transmitted symbols which can be expressed by,

$$s_{ZF} = \underbrace{(H^H.H)^{-1}.H^H}_{\text{Moore−Penrose pseudo−inverse}}.x \tag{2}$$

The channel matrix $H$ is $QR$ decomposed into two parts as$H = QR$.

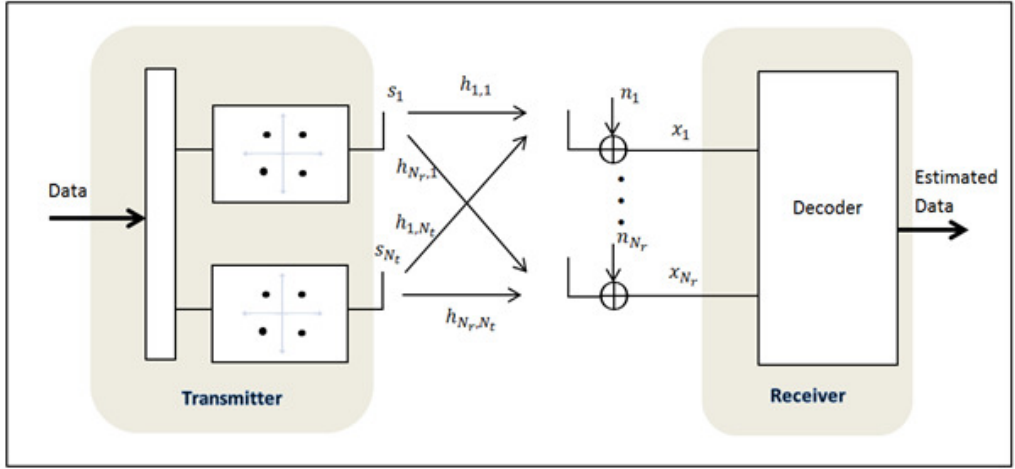

Figure 1. MIMO system with $N_t$ Transmitter and $N_r$ Receiver antennas.

## 3. LATTICE REDUCTION TECHNIQUE

We can interpret the columns $h_l$ of the channel matrix $H$ as the basis of a lattice and assume that the possible transmit vectors are given by $\mathbb{Z}^m$, the m dimensional infinite integer space. Consequently, the set of all possible undisturbed received signals is given by the lattice.

$$L(H) = L(h_1, \ldots, h_m) := \{\sum_{l=1}^{m} h_l \, | h_l \in \mathbb{Z}\} \tag{3}$$

The LR algorithm generates a lattices reduced and near-orthogonal channel matrix $\widetilde{H} = H.T$. With matrix $\widetilde{H} = H.T$ generates the same lattice as $H$, if and only if the m × m matrix T is unimodular [6], i.e. T contains only integer entries and $det(T) = \pm 1$:

$$L(\widetilde{H}) = L(H) \Leftrightarrow \widetilde{H} = HT \, and \, T \, is \, unimodular \tag{4}$$

Also,

$$\widetilde{H}.T^{-1} = H \tag{5}$$

We can find multiple bases that can be included in the space $L$, and the goal of the LR algorithm is to find a set of least correlated base with the shortest basis vectors [2].Initially, an efficient (but supposed not optimal) way to determine a reduced basis was proposed by Lenstra, Lenstra and Lovàsz [1].Where they defined (LLL-Reduced): A basis $\widetilde{H}$ with QR decomposition $\widetilde{H} = \widetilde{Q}.\widetilde{R}$ is called LLL-reduced with parameter δ with $(1/4 < \delta \leq 1)$, if

$$\left|\widetilde{R}_{i,j}\right| \leq \tfrac{1}{2}.\left|\widetilde{R}_{i,i}\right| \, for \, 1 \leq i < j \leq m \tag{6}$$

And

$$\delta\left|\widetilde{R}_{j-1,j-1}\right|^2 \leq \left|\widetilde{R}_{j,j}\right|^2 + \left|\widetilde{R}_{j-1,j}\right|^2 \, for \, j = 2, \ldots, m \tag{7}$$

The first condition is called, size-reduced and the second one is called Lovàsz condition. The parameter δ plays an important role to the quality of the reduced basis. We will assume $\delta = 3/4$ as proposed in [1]. After applying the QR decomposition of H, doing successive size-reduces if the condition is fulfilled, the algorithm exchanges two vectors if Lovàsz condition is not fulfilled to generate $T$, compute $\tilde{R}$ and $\tilde{Q}$. And so, the LLL algorithm will output $\tilde{Q}$, $\tilde{R}$ and $T$.

Looking to the LLL algorithm [1], one important element of its complexity is related to the fact that the LLL algorithm is applied for the real integer vectors, it is mandatory to reformulate the different matrices to their real-valued form, so we got:

$$H^{real} = \begin{bmatrix} Real(H) & -Im(H) \\ Im(H) & real(H) \end{bmatrix} \tag{8}$$

$$x = \begin{bmatrix} Real(x) \\ Im(x) \end{bmatrix} \tag{9}$$

$$s = \begin{bmatrix} Real(s) \\ Im(s) \end{bmatrix} \text{ and } n = \begin{bmatrix} Real(n) \\ Im(n) \end{bmatrix} \tag{10}$$

This kind of reformulation increases the number of operations and adds more latency for the system.

The idea behind LR-aided linear detection is to consider the equivalent system model and perform the nonlinear quantisation on it [7]. In fact, if we combine equations (1) and (5), we can get:

$$x = \tilde{H} . \underbrace{T^{-1}.s}_{z} + n \tag{11}$$

With $z = T^{-1}.s$ the equivalent model and in this case $\tilde{H}$ will represent a better channel quality. And so, the detector can be represented with an equivalent model with better performance due to the less noise enhancement increased by $\tilde{H}$. Thus, the basic idea behind approximate lattice decoding (LD) is to use LR in conjunction with traditional low-complexity decoders. With LR, the basis B is transformed into a new basis consisting of roughly orthogonal vectors [8].

After processing the Zero Forcing lattice reduction (ZF-LR) mechanism and by combining equations (2) and (11), we can generate:

$$\tilde{z}_{ZF-LR} = T^{-1}.\tilde{s}_{ZF} = \tilde{H}.x = z + \tilde{H}.n \tag{12}$$

The complex form of this algorithm was presented by Gan and Mow in [3]. But we can clearly identify that this extension keeps the excessive number of iteration and also add more computation latency by introducing the real and imaginary elements in the different conditions of the algorithm. For this reasons, Vetter proposed another variety of the complex LLL, than Ling [7] proposed a fixed complexity LLL (FCLLL). As recapitulation the modifications for the LLL algorithm was for three points:

- Avoid the complex to real vector transformations (reduce the number of loops).
- The reduction of the number of the LLL iteration to a fixed number.
- The use of a flag to track column exchanges. When no column swap happens, the FCLLL ends with an LLL reduced basis.

The different enhancements for the original algorithm where looking for limited iterations in term of stopping criteria, like in [2] and [5].

In next section we will consider the proposal in [2] and start form the Wen's algorithm as described in table 1 where, Wen proposed an enhanced form of Vetter's algorithm. The proposal is based on an improved column traverse strategy and an enhanced termination criterion for practical LR-aided SIC MIMO detection.

Table 1.  The Fixed Complex LLL algorithm [2]

| | |
|---|---|
| **Input:** | $H$: the channel complex matrix |
| **Output:** | $\tilde{R}, \tilde{Q}, T$ |
| 1 | Initialization: $T = I_{N_t}; N_t; \; CSflag = ones(1, N_t + 1); kSeq; N_{Max};$ |
| 2 | $[Q, R] := qr(H);$ |
| 3 | $\delta = 3/4$ |
| 4 | $n_{iter} = 1$ |
| 5 | $kSeq_{idx} = 1$ |
| 6 | $while \; (n_{iter} \leq N_{Max}) \&\& \; (sum(CSflag(2:1:N_t)) \neq 0$ |
| 7 | $k = kSeq(kSeq_{idx}) \; ;$ |
| 8 | $CSflag(k) = 0;$ |
| 9 | $for \; k = 2:N_t$ |
| 10 | $\mu := (\tilde{R}(l, k)/\tilde{R}(l, k))$ |
| 11 | $if \; \mu \neq 0$ |
| 12 | $\tilde{R}(1:l, k) := \tilde{R}(1:l, k) - \mu \, . \, \tilde{R}(1:l, l)$ |
| 13 | $T(:, k) := T(:, k) - \mu \, . \, T(:, l)$ |
| 14 | $end$ |
| 15 | $end$ |
| 16 | $if \; \delta . \tilde{R}(k-1, k-1)^2 > \tilde{R}(k, k)^2 + \tilde{R}(k-1, k)^2$ |
| 17 | $k \; to \; k-1 \; columns \; swap \; for \; \tilde{R} \; and \; T$ |
| 18 | $Computing \; the \; \Theta \; Matrix:$ $$\Theta = \begin{bmatrix} \bar{\alpha} & \bar{\beta} \\ -\beta & \alpha \end{bmatrix} with \quad \begin{aligned} \alpha &= \frac{\tilde{R}(k-1, k-1)}{\left\| \tilde{R}(k-1:k, k-1) \right\|} \\ \beta &= \frac{\tilde{R}(k, k-1)}{\left\| \tilde{R}(k-1:k, k-1) \right\|} \end{aligned}$$ |
| 19 | $\tilde{R}(k-1:k, k-1:m) := \Theta . \tilde{R}(k-1:k, k-1:m)$ |
| 20 | $\tilde{Q}(:, k-1:k) := \tilde{Q}(:, k-1:k) . \Theta^T$ |
| 21 | $CSflag(k-1:1:k+1) = 1;$ |
| 22 | $end$ |
| 23 | $end$ |
| 24 | $n_{iter} = n_{iter} + 1$ |
| 25 | $kSeq_{idx} = kSeq_{idx} + 1$ |
| 26 | $end$ |

# 4.MODIFIED ALGORITHM AND STUDY OF THE EFFECT OF MAX ITERATION ON THE FCLL ALGORITHM

In this section we will start from the FCLLL algorithm proposed by Wen [2] and we will try to do modify it. In fact, for line 21 there is the table " $CSflag$ " which is a condition for the loop as mentioned in line 6. The summation of the elements of this "table" seems to add $N_t - 1$ more addition operations that need to be computed for each loop. So, for us, it will be better to come back to the single element condition as mentioned in [9]. A second remark, the Lovàsz condition such as described in line 16, is representing four complex multiplication, one addition operation and one subtraction operation (which can be considered as addition operation). All of them are complex and being running in loop. It will be better to use the Siegel condition which is always fulfilling the Lovàsz condition and we can go more to show that it reduce the computing operations [10] & [11]. The representation is below:

$$\left|\tilde{R}_{j-1,j-1}\right|^2 \leq \zeta \left|\tilde{R}_{j,j}\right|^2 \tag{13}$$

Where $\zeta$ is chosen from $[2, 4]$,

To have the Siegel condition fulfilled, we have to check if:

$$\frac{1}{\zeta} \cdot \left|\tilde{R}_{j-1,j-1}\right|^2 > \left|\tilde{R}_{j,j}\right|^2 \tag{14}$$

Whatever the value of $\zeta$, 2 or 4, we can get:

$$\delta \cdot \left|\tilde{R}_{j-1,j-1}\right|^2 > \left|\tilde{R}_{j,j}\right|^2 \tag{15}$$

Since $\delta > \frac{1}{\zeta}$,

So, we can modify the algorithm in table 1 to get the below one as described in table 2.

Table 2: The proposed Modified Complex LLL algorithm

| | |
|---|---|
| **Input:** | $H$: the channel complex matrix |
| **Output:** | $\tilde{R}, \tilde{Q}, T$ |
| 1 | Initialization: $T = I_{N_t}$; $N_t = size(H, 2)$; $CSFlag = 0$; $IterMax = 6$; |
| 2 | $[Q, R] := qr(H)$; |
| 3 | $\delta = 3/4$ |
| 4 | $Iter = 0$ |
| 5 | $while \ (CSflag == 0) \ \&\& \ (Iter \leq IterMax)$ |
| 6 | $CSflag = 1$; |
| 7 | $Iter = Iter + 1$; |
| 8 | $for \ k = 2: N_t$ |
| 9 | $\mu := round(\tilde{R}(l, k)/\tilde{R}(l, k))$ |
| 10 | $if \ \mu \geq 1$ |
| 11 | $\tilde{R}(1:l, k) := \tilde{R}(1:l, k) - \mu \cdot \tilde{R}(1:l, l)$ |

12          $T(:,k) := T(:,k) - \mu . T(:,l)$

13        $end$

14      $end$

15    $if\ \delta.\tilde{R}(k-1,k-1)^2 > \tilde{R}(k,k)^2$

16         $k\ to\ k-1\ columns\ swap\ for\ \tilde{R}\ and\ T$

           $Computing\ the\ \Theta\ Matrix$:

17     $\Theta = \begin{bmatrix} \bar{\alpha} & \bar{\beta} \\ -\beta & \alpha \end{bmatrix} with \begin{array}{l} \alpha = \dfrac{\tilde{R}(k-1,k-1)}{\lVert \tilde{R}(k-1:k,k-1) \rVert} \\[2mm] \beta = \dfrac{\tilde{R}(k,k-1)}{\lVert \tilde{R}(k-1:k,k-1) \rVert} \end{array}$

18     $\tilde{R}(k-1:k,k-1:m) := \Theta.\tilde{R}(k-1:k,k-1:m)$

19     $\tilde{Q}(:,k-1:k) := \tilde{Q}(:,k-1:k).\Theta^T$

20    $CSflag =0;$

21    $end$

22     $k := k+1$

23    $end$

24   $end$

In the proposed algorithm we have modified the line 5 by avoiding "CSflag" table summation presented in table 1 and proposed in [2]. This will help to reduce additional processing operations which will help to "relax" the algorithm in term of complexity and decoding timing. In fact, and as described in the previous section, the contribution of the elementof this "table" in the algorithm doesn't exceed the termination condition. The importance of this modification can be observed in the next sections, especially of the gain in terms of complexity.

We can clearly observe that the max iteration number ($N_{Max}$) is a condition to exit from the loop and so, it can increase the number of computation operations. This means, there is an ideal max iteration value that above it the system becomes exponential complex without large BER enhancement. Also, the modifications in line 10 and line 15 will help to reduce the processing operations because the algorithm will converge quicker than the initial version. In fact, the Siegel condition helps to relax the processing operations like presented in [10].

So, it's interesting to evaluate the effect of the Max iteration on the BER performance and also the system complexity. For this, we tried to do the simulation of the algorithm with varying the value of the Max iteration.

# 5. SIMULATION RESULTS AND EFFECT OF THE MAX ITERATION ON THE FCLLL ALGORITHM

For our simulation, we will consider the 16QAM constellation, ZF equalization will be checked. The MIMO model will be $4 \times 4$ ; means a 4 antennas at both transmitter and receiver side. We used a frame size of $10^5$ . We will indicate inline any changes to the above configuration. In the flowing figures, we tried to increase the max iteration number from 4 to 18.

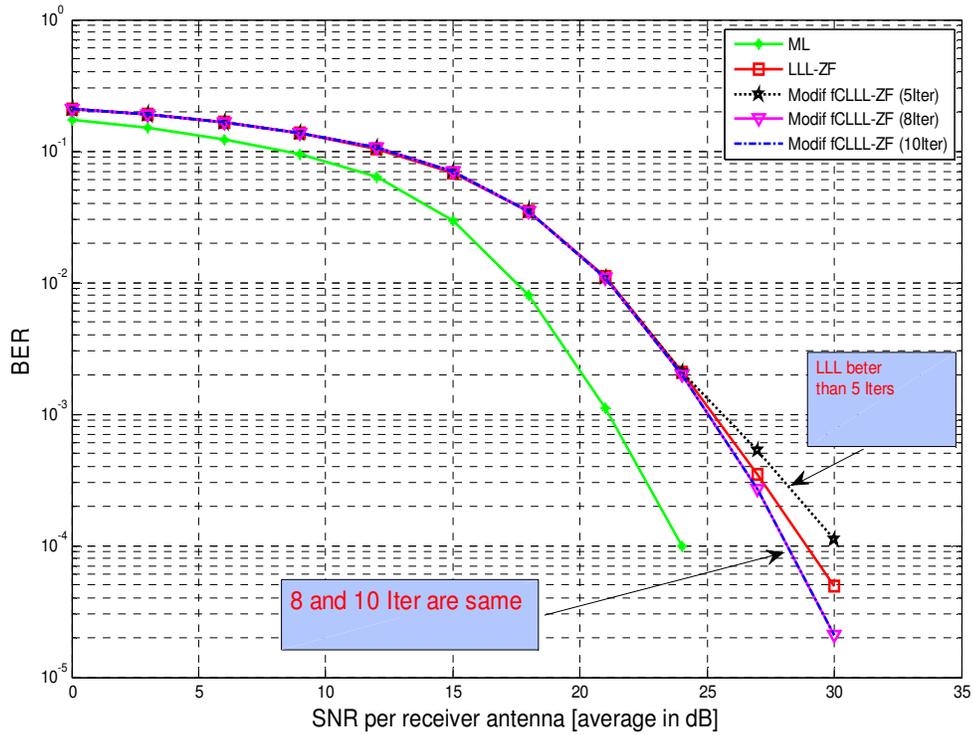

Figure 2: $4 \times 4$ Modified Complex LLL with 16QAM and ZF Bit error rate results. 5, 8 and 10 IterMax results compared to ML and ordinary LLL.

Observing figure 2, the ML curve is outperforming all different curves. But we should note that the ML scheme is extremely complex to implement. So, we are indicating it just for reference and comparison reasons.Another quick remark is that comparing the FCLLL curves and the ordinary LLL algorithm we can see that for IterMax ≤ 5, the LLL is better comparing FCLLL. But for MaxIter equal to 5, the two curves are overlapping till SNR equal to 24dB and after the deviation is minimal. Which means that in terms of performances we are still in an acceptable range and so it will be interesting to push the analysis and also evaluate the gain in complexity and processing operations.

In the figure 3, we increase the max iteration value from 4 to 18 to observe if any threshold value for this parameter; that allow to reach a better result with lower iterations.

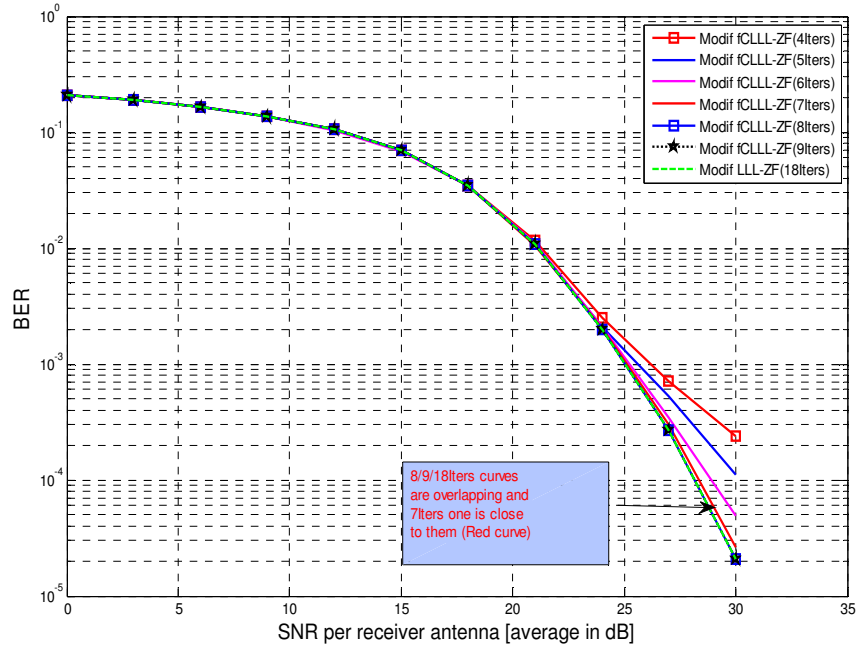

Figure 3: BER results by varying the IterMax: curves for IterMax equal to 4,5,6,7, 8,9 and 18.

Looking to figure 3 we can observe that, staring from $IterMax \geq 6$, the FCLLL become similar or better than the LLL. But, starting from 8 iteration we can observe that no improvement for the BER.

Means, the curves remain overlapping each other. This leads us to conclude that no need to increase the $IterMax$ parameter above 8 iterations. Else, the system became costly comparing to its performance. For $IterMax$ between 5 and 8, we can push the analysis. In fact, the LLL algorithm as described in [1], will do a minimum of $2.N_t$ loops; taking in consideration the fact that the size of the channel real-valued matrix H used for the LLL algorithm is double of the complex matrix H. Thus, for this case and with 8 IterMax we are exactly in the same condition as the LLL algorithm. From another point of view, a $IterMax \leq 4$ will show a BER degradation. This is related to the fact that the algorithm will do a column swap for only half of the possible columns of the matrix. If we consider 5, 6 and 7 as IterMax we can see that we more or less close to the LLL algorithm, since the difference is observed only for the high SNR and the deviation is minimal. In the case of $IterMax = 6$ the BER curve is almost overlapping the ordinary LLL curve. From our point of view, using the $IterMax$ equal to 6 seems to be the recommended value, since it has a good complexity to performance balance.

Figures 4 and 5 show a zoom on the different curves to illustrate our analysis.

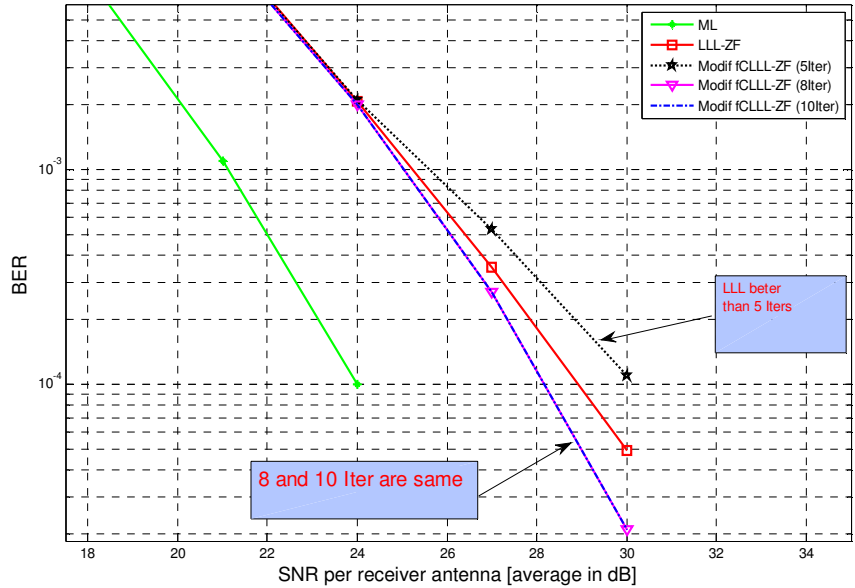

Figure 4: 4 × 4 Modified Complex LLL with 16QAM and ZF Bit error rate results. 5, 8 and 10 IterMax Zoomed curves compared to ML and ordinary LLL.

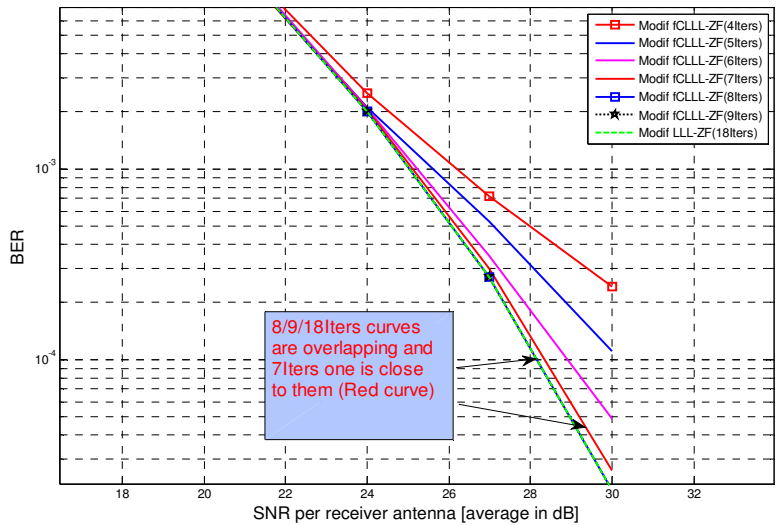

Figure 5: BER results by varying the IterMax: Zoomed curves for IterMax equal to 4,5,6,7, 8,9 and 18.

We remark that starting from $IterMax = 8$ the system BER performance reach the saturation but also the system computing operation are increasing according to the $IterMax$. Means, the complexity continue increasing function of $IterMax$ but the BER performance will saturate. Just looking to numbers, the BER saturation is reached for 8 $IterMax$ and the BER performance for 6 $IterMax$ is same as the ordinary LLL. So, we got same performances as ordinary LLL with a gain of ¼ of operations. It's a good performance vs complexity balance to be considered...

# 6. COMPLEXITY ANALYSIS

In this section we will discuss the complexity aspect of our proposal and show the profits and benefits of our proposal.

First we will give some details about the operation done by the algorithm. In [12] it was presented that a real matrix multiplication of A ($N \times M$) and B ($K \times M$) leads to matrix C ($N \times M$) and the overall operations are $N(K-1)M$ addition operation and $NKM$ multiplication operations. It is also known that a complex addition is equivalent to two real addition operations. In fact, for the complex case we will add the real and imaginary parts separately. For the complex multiplication it is different and the operation can be written as below:

$$(a+ib)*(c+id) = (ac-bd) + i(bc+ad)$$

$$= (ac-bd) + j\big((a+b)(c+d) - ac - bd\big) \tag{16}$$

The first options can be done in four multiplications and two additions (assuming that a subtraction can done via an addition operation). The second option can be done in three multiplications and five additions. But the first option is almost used. So, we will consider it. Also, in [13] it was shown that the different arithmetical operation requires different FLOPS. In the table below we present the number of FLOPS needs for each operation (for real values) [13].

Table 3: FLOPS vs operations

| Operation | Add | Mult | Sqrt | Div |
|---|---|---|---|---|
| **Nombre of FLOPS** | 1 | 1 | 8 | 8 |

- The size reduction require $(N_{Max} - 2) \times \{1 \times (Div) + 2 \times (Mult + Add)\}$
- The lovàsz condition require $\{4 \times (Mult) + 2 \times (Add)\}$
- Colum swap require $N_r \times \{3 \times (Add)\}$
- The Givens rotation matrix computation require $\{2 \times (Div) + 2 \times (Mult) + 1 \times (Add) + 1 \times (Squrt)\}$
- The rotation operation for R (matrix multiplication) require $2 \times \{2 \times (Mult) + 1 \times (Add)\} \times N_r$
- The rotation operation for Q (matrix multiplication) require $2 \times \{2 \times (Mult) + 1 \times (Add)\} \times 2$
- The CSflag condition sum require $N_t(Add)$

Also, the complex division and square root operations consists of many real operations.

- A square root of complex value require $\{1 \times (Div) + 3 \times (Mult) + 2 \times (Add) + 3 \times (Sqrt)\}$ of real values.
- A complex value division require $\{1 \times (Div) + 8 \times (Mult) + 4 \times (Add)\}$

All these operation will be running in loop for $N_{Max}$ iterations for MIMO $8 \times 8$ scheme.

Table 4: Complexity gain

| | FLOPS | Performance | Comments |
|---|---|---|---|
| LLL algorithm $N_{Max} = \infty$ | 12200 < | | |
| Wen's algorithm [5] $N_{Max} = 18$ | < 9300 | Outperform LLL (11dB at $10^{-4}$) | |
| Our algorithm $N_{Max} = 18$ | < 9100 | Gain 2dB at $10^{-4}$ vs LLL | The most important point is that we reach same performance with ~31% of FLOPS gain |
| Our algorithm $N_{Max} = 8$ | < 6200 | Gain 2dB at $10^{-4}$ vs LLL | |
| Our algorithm $N_{Max} = 6$ | < 5800 | Loose 2dB at $10^{-4}$ vs LLL | Gain ~36% of FLOPS and the performance degradation is minimal (2dB) |

The table above shows that with our proposal we can reach approximately the same performances as the LLL algorithm with reducing 36% of FLOPS. This is important in term of decoding delay. In fact, we can avoid some decoding delay and achieve the same performance with limited iteration number ($N_{Max} = 8$).

## 7. CONCLUSIONS

In this paper we proposed some modifications to the FcLLL algorithm proposed by Wen [2]. Simulation results show that for $4 \times 4$ MIMO system, there is min and max values for the $IterMax$ (5 to 8) where the BER performances seems to be good (more or less near to the original LLL results) and also the system complexity remains reasonable. Outside these limits the complexity vs performance balance become undesirable. And the extra iterations don't enhance the performance. Thus, to implement this algorithm we recommend an ideal value of $IterMax = 6$ which allows having a BER quite same as the original LLL and limits the iterations loop. In fact, with this recommended value we can gain ~36% of operations and the BER degradation will be ~2dB at $10^{-4}$. The challenge of our proposal was to not bring many changes to the original algorithm, but to identify the possible points that we can enhance in order to relax the processing operations and complexity while keeping good performance results (nearest to the original algorithm). Such study and the presented results aim to help the industry using a low complexity, low cost and high performance solution based on the LLL decoding technique.

## REFERENCES

[1] A. K. Lenstra, H. W. Lenstra, and L. Lovàsz, (1982), "Factoring polynomials with rational coefficients," in Math. Ann, vol. 261, pp. 515 - 534.

[2] Qingsong Wen, Qi Zhou, and Xiaoli Ma, (2014), "An Enhanced Fixed-Complexity LLL Algorithm for MIMO Detection", Globecom 2014 - Signal Processing for Communications Symposium.

[3] Y. H. Gan and W. H. Mow, (Dec. 2005) "Complex lattice reduction algorithms for low-complexityMIMO detection," in Proc. IEEE Global Telecommun. Conf., St. Louis, MO, vol. 5, pp. 2953–2957.

[4] C. P. Schnorr and M. Euchner, (1994) "Lattice Basis Reduction: Improved Practical Alorithms and Solving Subset Sum Problems". Mathematical Programming, vol. 66, pp. 181.191.

[5] Z. Ma, B. Honary, P. Fan, and E. Larsson, (Jun. 2009), "Stopping criterion for complexity reduction of sphere decoding," IEEE Commun. Lett., vol. 13 , no. 6, pp. 402–404.

[6] D. Wubben, D. Seethaler, J. Jalden, and G. Matz, (April 2011), "Lattice reduction," in IEEE Signal Processing Magazine, vol. 28, no. 4, pp. 70 - 91.

[7]  C. Ling, W. H. Mow, and N. Howgrave-Graham, (Mar. 2013), "Reduced and Fixed- Complexity Variants of the LLL Algorithm for Communications," IEEE Trans. Commun., vol. 61, no. 3, pp. 1040–1050.

[8]  Md Hashem Ali Khan, Jin-Gyun Chung and Moon Ho Lee, (2015), "Lattice reduction aided with block diagonalization for multiuser MIMO systems", EURASIP Journal on Wireless Communications and Networking 2015:254, DOI 10.1186/s13638-015-0476-1

[9]  H. Vetter, V. Ponnampalam, M. Sandell, and P. A. Hoeher, (Apr. 2009), "Fixed complexity LLL algorithm," IEEE Trans. Signal Process., vol. 57, no. 4, pp. 1634–1637.

[10] B. Gestner, W. Zhang, X. Ma, and D. V.Anderson, (Apr. 2011) ". Lattice Reduction for MIMO Detection: FromTheoretical Analysis to Hardware Realization," IEEE Trans. Circuits Syst. I, Reg. Papers, vol. 58, no. 4, pp. 1549-8328.

[11] L. G. Barbero, D. L. Milliner, T. Ratnarajah, J. R. Barry, and C. F. N. Cowan, (Jun. 2009), "Rapid prototyping of Clarkson's lattice reduction for MIMO detection," in Proc. IEEE Int. Conf. Commun., Dresden, Germany, pp. 1–5.

[12] Markus Bläser, (2013), "Fast Matrix Multiplication", Theory of Computing, Graduate Surveys , vol. 5, p. 1-60

[13] Ameer Youssef , Mahdi Shabany , P. Glenn Gulak, (May 2011), "Performance analysis of lattice-reduction algorithms for a novel LR-compatible K-Best MIMO detector," Conference: International Symposium on Circuits and Systems, DOI: 10.1109/ISCAS.2011.5937662, Rio de Janeiro, Brazil.